\documentclass[aps,prb,twocolumn,superscriptaddress,showpacs]{revtex4}
\usepackage{graphicx}

\def\be{\begin{equation}}
\def\ee{\end{equation}}
\def\ba{\begin{eqnarray}}
\def\ea{\end{eqnarray}}

\def\bc{\begin{center}}
\def\ec{\end{center}}

\begin{document}

\title{Long-range nature of surface-enhanced Raman scattering}

\author{V.~I.~Kukushkin}
\affiliation{Institute of Solid State Physics, RAS, Chernogolovka
142432, Russia}

\author{A.~B.~Van'kov}
\affiliation{Institute of Solid State Physics, RAS, Chernogolovka
142432, Russia}

\author{I.~V.~Kukushkin}
\affiliation{Institute of Solid State Physics, RAS, Chernogolovka
142432, Russia}

\date{\today}

\begin{abstract}
The long-range action of surface-enhanced Raman scattering (SERS)
is probed via distance-dependent measurements of molecular Raman
spectra. To this end, identical SERS substrates composed of
irregular silver nanoisland arrays were covered by dielectric
spacer layers with variable thickness, and the strength of the
SERS signal produced from analyte molecules deposited on top of
the structure was analyzed.  The obtained distance dependence of
the signal strength exhibited a shelf-like behavior up to 30\,nm
away from the enhancing surface and then rapidly decreased further
away. Thus, the observed behavior of the electromagnetic mechanism
of SERS enhancement in metal island films contradicts the widely
accepted picture of extremely rapid (2--3\,nm) decay of
SERS-enhancement of 2D nanoparticle ensembles. Because of the
observed steady enhancement factors at distances of $\sim$30\,nm
from the surface, SERS can be used for probing the spectra of
macromolecules or other objects relatively distant from the metal
surface.
\end{abstract}


\maketitle

Since its first observation by Fleischmann {\it et
al.}(1974)~\cite{Fleischmann74} surface-enhanced Raman scattering
(SERS) has been thoroughly investigated as an amazing physical
phenomenon itself and one of the most promising tools for
analytical applications. The first perception of SERS as a giant
enhancement over conventional Raman scattering in experiments of
Jeanmaire, Van Duyne~\cite{VanDuyne} and Albrecht,
Creighton~\cite{Albrecht} was followed by an extensive theoretical
and experimental analysis, searching for the most general
explanation of the enhancement mechanism. The enhancement of Raman
scattering signals from organic molecules absorbed on
nanostructured metal surfaces and photoexcited in a certain
spectral range has been shown to reach 6--10 orders of
magnitude~\cite{VanDuyne,Kneipp1984}. Further enhancement of up to
$\sim$14 orders of magnitude was observed on molecules residing in
silver colloidal aggregates, enabling single molecule
detection~\cite{Kneipp1997}.

It is now generally agreed that more than one effect contributes
to the total enhancement of Raman signals. The enhancement
mechanisms are roughly divided into so-called electromagnetic (EM)
field enhancement~\cite{GerstNitzan,Platzman80,Wang80} and
chemical first-layer effects~\cite{Moskov85,Otto84,Otto92}. The
electromagnetic enhancement is caused by the enhanced local
optical fields at the place of the molecule nearby the metal
surface due to excitation of electromagnetic resonances, called
surface plasmon polaritons. For isolated silver or gold spheroidal
nanoparticles typical values for electromagnetic enhancement of
SERS are on the order of~\cite{Kerker80,Kerker81} $10^6-10^7$.

For the more sophisticated case of closely spaced interacting
particles (or clusters of particles), the individual dipole
oscillators of the small particles couple, thereby generating
normal modes of plasmon excitation that embrace the cluster.
According to theoretical evaluations, the excitation is not
distributed uniformly over the cluster but tends to be spatially
localized in so-called "hot"
areas~\cite{Shalaev96,Sarychev98,Moskov99}.

Effects of chemical enhancement arise from the electronic coupling
between the adsorbate molecule and the metal surface and
excitation of adsorbate-localized electronic resonances (or charge
transfer resonances). The chemical enhancement is valid solely for
molecules in direct contact with the metal surface, and hence, it
is referred to as a first-layer enhancement. The magnitude of
chemical enhancement has been estimated to reach no more than two
orders of magnitude~\cite{Otto84}. The combined action of
electromagnetic and chemical enhancement mechanisms is referred to
as surface-enhanced resonant Raman scattering with multiplied
enhancement coefficients.

Although both mechanisms combine their enhancements to reach
better signal gains, probing them separately is crucial for
understanding their physics. Electromagnetic enhancement is
universal for all types of molecules and can be controlled and
optimized by specially designed SERS-active nanostructures.

For applications of SERS in ultrasensitive detection, not only the
values of the enhancement factor, but the range of action as well
are key aspects. In the literature there is still some ambiguity
concerning the last aspect of EM enhancement mechanism, what in a
certain extent restricts the applications of SERS in analytical
methodics.

In most previously published investigations of SERS distance
dependence, the Raman-active adsorbate was spaced from the surface
using self-assembled monolayers of long organic molecules
chemisorbed to the metal surface and thus the overall effect was
subject to interplay of chemical and EM enhancement mechanisms or
even suffered from modification of the metal surface electron
states. Several studies~\cite{Kennedy99,Compagnini99} performed on
long chains of alkanethiols attached to silver surfaces show that
the enhancement for remote chemical groups vanishes at extremely
short distances ($2-3$\,nm), which is in accordance with the
theoretically predicted power law\cite{Kennedy99} $\sim
\left(1+\frac{h}{r}\right)^{-10}$ derived for a chaotic array of
{\it non-interacting} nanoparticles, where $r$ is the mean radius
and $h$ is the molecule-to-surface distance. Experiments using
spacers created on nanostructured surfaces by atomic layer
deposition of Al$_2$O$_3$ also show very short-range SERS
enhancements~\cite{Faraday2006}, possibly resulting from the
modification of the metal surface properties during the atomic
layer deposition process. Thus, in the above-mentioned cases it is
unclear what type of SERS enhancement is probed -- chemical,
electromagnetic, or a combination of the two.

We present a study of SERS distance dependence that utilizes
chemically passive, uniform spacer layers with controlled
thickness. To this end, a silver-based, SERS-active surface is
covered by a transparent dielectric layer of variable thickness,
and analyte molecules are settled on top of the spacer. This
approach enables a distinctive study of the pure EM properties of
SERS enhancement, irrespective of the chemical enhancement
aspects.

{\bf Experimental}

The experimental samples were prepared using thermal vacuum
deposition. Polished silicon plates were used as wafers for
deposition of layered metal-dielectric structures. The sequence of
layers was as follows: (a)a relatively thick (50\,nm) silver
screening layer; (b)an insulating silicon monoxide (SiO) layer
with a thickness of ~15\,nm; (c) a SERS-active nanoparticle layer
formed by a silver island film with mass thickness of 6\,nm; and
(d) a SiO dielectric spacer layer with variable mass thickness,
ranging from 0 to 60\,nm. The final passivating layer was intended
to separate the analyte molecules from the metal nanoparticles,
and it was the primary parameter of the examination. The
morphology of the first three layers in the SERS substrates and
the deposition parameters (rate, pressure, and temperature)  were
fixed throughout the experiments. The heating power was kept
constant to provide deposition rates of ~0.5\,\AA/s for Ag, and
~1.5\,\AA/s for SiO. The deposition pressure was lower than
$5\times10^{-6}$\,torr. The distance dependence of SERS
enhancement was studied by varying the thickness of the
passivating dielectric layer in different samples. The set of
samples consisted of SERS substrates with passivating layer
thicknesses varying from 0 to 60 nm, with increments of 10 nm.
Additionally, the dependence was studied on the samples with
continuously varying thickness of the passivating layer across the
surface. These gradient samples benefit from guaranteed
homogeneity of the SERS-active layers across the sample surface,
with only one parameter being varied -- the thickness of SiO
spacer. To produce such gradient substrates during the single-load
of a vacuum deposition camera, a specially shaped mask was
utilized (See bottom inset of Fig.1). The basic layers (a),(b),
and (c) were deposited while the opening in the mask was precisely
against the substrate surface, allowing the particle beams to
settle homogeneously onto the surface. After the SERS-active layer
was deposited, the axial-symmetric mask was set to rotation with
rate 15\,r/min, and the time-averaged exposure for the passivating
layer became a variable function of the radial coordinate in
accordance with the mask profile.

The surface morphology of the samples was studied using an
electron beam microscope (EBM) Jeol model JSM 7001F with a
resolution of ~1\,nm. The gradient profile of the passivating
layer across the substrate was defined by EBM analysis of the
sample cross section. The gradient SiO layer was sandwiched
between two 50-nm-thick silver layers, and the cross section of
this structure was analyzed and evaluated by the EBM. The
thickness calibration curve of the gradient profile was plotted in
coordinates (radial displacement--thickness). Independently, the
thickness calibration was evaluated numerically based on the shape
of the mask, and both calibration curves are plotted in Fig.1.
Furthermore, EBM measurements were important for ascertaining that
the SiO spacer layer is pinhole free and that it does not affect
the characteristics of measured distance dependence. For this
purpose, an additional 50-nm-thick layer of Ag was deposited on
top of the passivating SiO layer to obtain an EBM-image with
better contrast (See bottom-left inset of Fig.1). By analyzing EBM
images taken at different spacer thicknesses, we determined that
the spacer layer is pinhole free at spacer thicknesses greater
than 15\,nm.

\begin{figure}[h]\begin{center}
\includegraphics*[width=.5\textwidth]{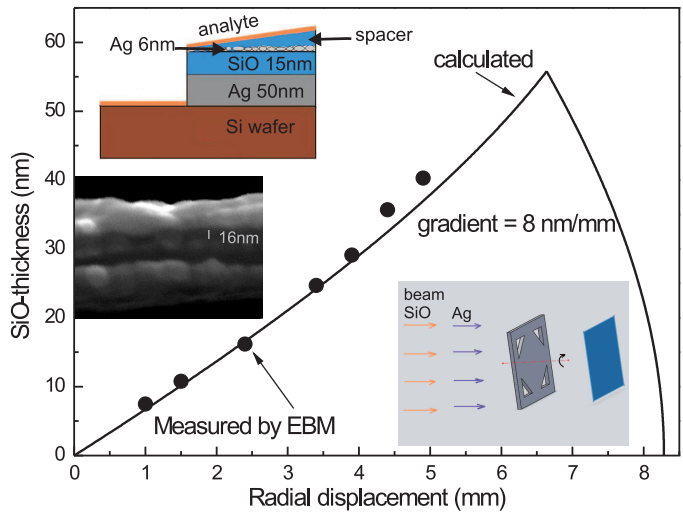}
\end{center}
\vspace{-2.mm} \caption{The calibration curve of a gradient
profile used for deposition of a passivating SiO layer. The smooth
curve represents the result of a numerical evaluation based on the
mask shape. The symbols are extracted from EBM imaging of the
cross section of layered structure. The top inset illustrates the
morphology of the SERS substrate with a metal island film covered
by a dielectric layer with gradually increasing thickness. The
bottom-right inset is an illustration of the vacuum deposition
process with a rotating mask. The bottom-left inset shows the EBM
image of cross section of the substrate covered by the 16\,nm SiO
layer and 50\,nm Ag layer. } \label{f.1} \vspace{-1.5mm}
\end{figure}

Different types of analytes were used to probe SERS enhancement:
Rhodamine 6G (R6G) and $\beta$-carotene dyes, and colorless
adenosine. Rhodamine 6G and $\beta$-carotene were deposited on
substrates by spin-coating 10\,$\mu L$ drops of $1-\mu M$ and
$10-\mu M$ ethanol solutions at 4000\,r/min until drying was
complete. Adenosine was dried out from a 2\,$\mu L$ drop of
$200\,\mu M$ water solution. These representatives were chosen
because of their qualitatively different properties as
Raman-active agents: both dye molecules have strong Raman
activity, but differ in fluorescence activity -- Rhodamine 6G is
fluorescent under 532\,nm excitation as opposed to
$\beta-$carotene at the same excitation. Colorless adenosine
exhibits moderate Raman activity and is non-fluorescent. The
distance dependence of SERS-enhancement on different types of
molecules was tested in order to achieve a general view on whether
resonant Raman enhancement plays a role in this effect.

Raman scattering experiments were performed at room temperature
using the EnSpectr micro-Raman spectrometer with $\lambda_{exc} =$
532\,nm excitation. The laser ($P = 10$ to $200\,\mu W$) was
focused onto the sample by a 20X microscope objective. The laser
spot, sized ~$5\,\mu m$, determined the spatial resolution of the
surface-scanning method. The scanning was performed using the
motorized XYZ-microstage M\"arzh\"auser with a resolution of
0.1\,$\mu m$. The exposure time for each spectrum was set as 1 s.

{\bf Results and Discussion}

Typical SERS spectra of analyte molecules are shown in Figs.2 and
3. The data shown in Fig.2 for $\beta-$carotene and adenosine are
obtained for the series of substrates with nominally fixed
morphology of the SERS-active nanostructure and with different
thicknesses of the SiO spacer layer. The enhancement coefficient
of nonpassivated SERS substrates was estimated to be between
$\sim5\times10^5$ and $10^6$, typical for such silver
nanostructures. The homogeneity of each substrate was investigated
via {\it x-y} scanning of the signal. On length scales of 2x2\,mm,
the rms deviation of the signal was found to be less than 3\%,
including deviations in the enhancement coefficient and in the
surface distribution of the deposited molecules. A waterfall
spectra (Fig.2A) explicitly displays the change in signal with
increasing molecule-to-nanoparticle layer separation. The inset
shows the dependence of the intensity of the strongest line with a
Raman shift of 1525\,cm$^{-1}$ on the spacer thickness. The
dependence is rather weak below the 30\,nm separation, after which
an abrupt collapse of signal is observed. Similar behavior is
observed in the series of substrates with another analyte molecule
-- adenosine-- applied using the $200-\mu M$ water solution
(Fig.2B). Here, the intensity of the 735\,cm$^{-1}$ line was
analyzed. Despite the common trend of the distance dependencies
that is obvious from these data, some inconveniences are
unavoidable in this scheme of experiments. They include
fluctuations in the output parameters of the substrates for each
process of vacuum deposition, fluctuations in the analyte molecule
deposition conditions, and the few points in this dependence.

\begin{figure}[h]\begin{center} \vspace{2.mm}
\includegraphics*[width=.5\textwidth]{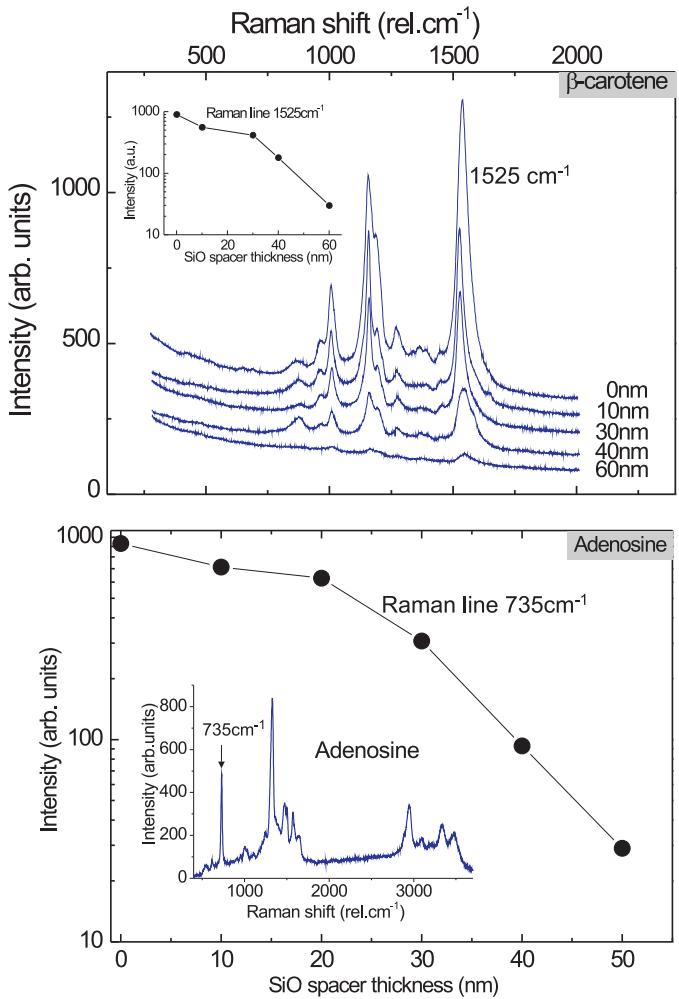}
\end{center}
\caption{(A) SERS-spectra of $\beta$-carotene molecules applied
over the SiO spacer layer with thickness from 0 to 60\,nm. The
inset shows the plot of the 1525cm$^{-1}$  Raman line intensity as
a function of spacer thickness. The molecules were deposited via
spin-coating of 10$\mu L$ drop of 10$\mu$M ethanol solution.
Excitation power is 200\,$\mu W$ at 532\,nm and exposure time is 1
s. (B) Plot of the SERS signal distance dependence for the
735cm$^{-1}$ Raman line of adenosine molecules, deposited from the
2$\mu L$ drop of $200-\mu M$ water solution. Discrete points
correspond to the signal obtained for substrates with SiO spacer
thickness varying from 0 to 50 nm. The inset shows a typical SERS
spectrum of adenosine.} \vspace{-1.5mm}
\end{figure}

Therefore, we prepared and analyzed another type of sample, having SERS substrates
with the spacer layer thickness increasing in a smooth gradient. Considering the characteristic length
scale of the distance dependence, the parameters for vacuum
deposition of the SiO spacer layer were chosen such that spacer thickness gradually increased from 0 to $\sim$50\,nm, with a
surface span of 8\,mm. The gradient samples were spin-coated using solutions of Rhodamine 6G and $\beta-$carotene molecules. The
spectra were recorded and analyzed across the gradient sector of
the sample surface such that the integral intensities
of several Raman lines were calculated and plotted as an
image plot shown in the inset of Fig.3A. The image plot represents the
distribution of the SERS signal from $\beta-$carotene on the
gradient sector of the SERS substrate. The distance dependence of
the 1525\,cm$^{-1}$ Raman line intensity was extracted from this
2D-array of data points along the bisecting line. The resulting
smooth dependence is shown in Fig.3A. It exhibits
shelf-like behavior from 0 to $\sim$25\,nm and
decreases rapidly at distances greater than 30\,nm. Similar
dependence was observed for R6G molecules, as shown in Fig.3B.

Note that the absolute values of SERS signal per molecule and per unit
power of optical excitation differ by orders of magnitude for the
different molecules used. For colorless adenosine molecules, the
SERS spectrum is the weakest, whereas for the R6G dye molecules with
absorption maximum at 523\,nm the SERS is resonant and the resulting cross section is stronger
by nearly three orders of magnitude. Irrespective of these
dramatic differences in SERS cross sections, all three types of
analyte molecules manifest similar distance-dependence behavior with a critical length parameter of approximately 25-30\,nm.

In essence the experiment proves that the applied method directly probes the
EM mechanism of SERS enhancement and excludes aspects such as chemical enhancement of the "first layer"
of molecules or any other resonant contributions to the signal. The
distance range of SERS enhancement should depend on the morphology
of SERS-active layers and reflect the surface plasmon localization
properties.

A length scale of 30\,nm substantially exceeds the mean radius of
nanoislands, and thus, the measured distance-dependence is in
marked contrast with the well-known power law $\sim
\left(1+\frac{h}{r}\right)^{-10}$ theoretically predicted for the
chaotic array of independent nanoparticles. However, the observed
shelf-like dependence indicates the existence of a new lengthscale
in metal nanoparticle ensembles. Since this length scale manifests
in the EM enhancement range, and the EM enhancement is in turn
caused by the field distribution of surface plasmon polaritons
(SPP), one may conclude that, firstly, SERS enhancement is
determined by the {\it collective} surface plasmon polaritons in
ensembles of neighboring nanoislands, and secondly, the work range
of SERS enhancement should be determined by the penetration depth
of the SPP electric field to the dielectric media.

In order to estimate this attenuation length SERS-active layers of
silver islands can be roughly modeled as a thin, flat metallic
layer sandwiched between two dielectric slabs. According to
classical electrodynamics, the penetration depth of the SPP electric
field in the dielectric is
$$\ell_p\sim\frac{\lambda_0}{2\pi}\frac{\sqrt{\left
|\varepsilon_m(\omega)+\varepsilon_d \right |}}{\varepsilon_d},$$
where $\lambda_0$ is the vacuum wavelength of exciting EM-field,
$\varepsilon_m(\omega)$ is the dielectric function of the metal,
and $\varepsilon_d$ is the average dielectric constant of the
surrounding dielectric. In the case of silver islands surrounded
by the 15\,nm SiO layer and air and wavelength
$\lambda_0\sim530$\,nm we substitute $\varepsilon_d\approx4$, and
from Ref.\onlinecite{Johnson72}, we use
$\varepsilon_m\approx-12+0.3i$ to obtain $\ell_p\approx60\,$nm.
Using these substitutions, we find that the length parameter for
the EM-field intensity, probed in optical experiments, is half as
long -- $\ell_{I}\approx\ell_p/2\approx30\,$nm -- which is in
reasonable agreement with measured values.

\begin{figure}[h]\begin{center} \vspace{2.mm}
\includegraphics*[width=.5\textwidth]{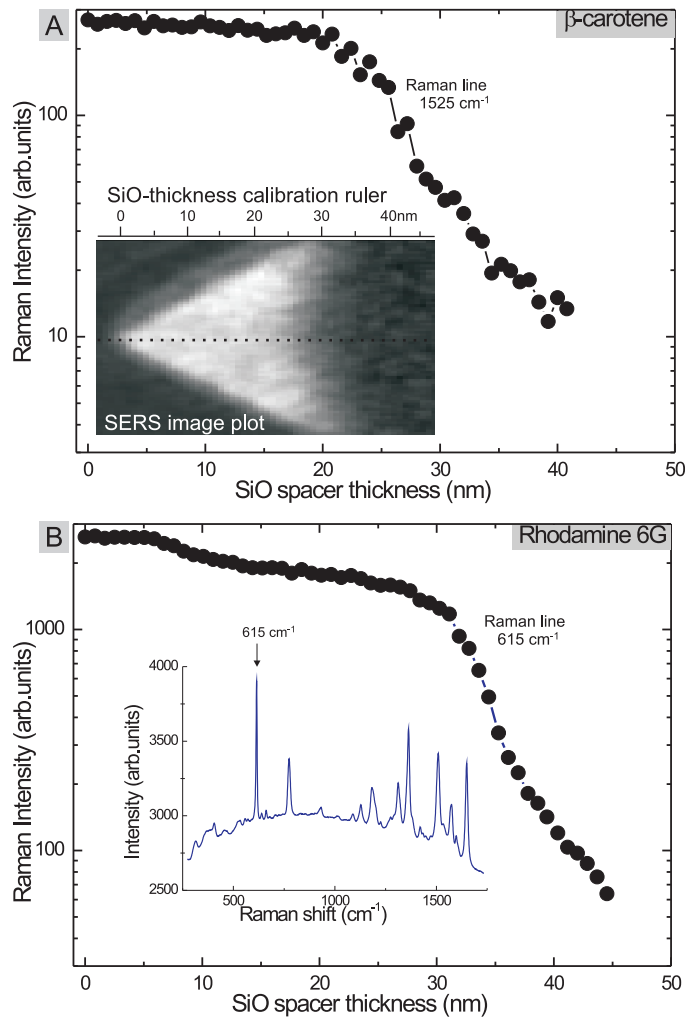}
\end{center}
\caption{SERS signal distance dependencies obtained for substrates
with gradually increasing SiO spacer thicknesses across the
surface. (A) Data for 1525cm$^{-1}$  Raman line of
$\beta$-carotene. The inset shows the image-plot of the Raman line
intensity over the substrate surface. The brightness of points
represents the level of Raman signal, and the triangle sector
corresponds to the SERS-active area of the sample with Raman
signal diminishing from the vertex to the base. The main plot is
extracted from data points along the triangle bisecting dotted
line. The upper ruler on the image displays the scale of spacer
thickness across the bisecting line. (B) Analogous continuous
distance dependence plot measured for Rhodamine-6G molecules with
SERS spectrum is shown in the inset. Excitation power is 10\,$\mu
W$ at 532\,nm and exposure time is 1 s.} \vspace{-1.5mm}
\end{figure}

{\bf Conclusion}

In conclusion, this work studied the distance dependence of
SERS enhancement of silver nano-island films.
The layered Ag/SiO SERS substrates with fixed morphology were
coated with an optically transparent passivating spacer layer composed of SiO, with
variable thickness, and the signal dependence on the thickness of the
spacer layer was studied. Different types of analyte molecules showed
qualitatively coinciding results. The SERS enhancement reduced
slightly (less than half) up to distances of $\sim$30\,nm, after which
it reduced abruptly. These results indicate that for such SERS substrates, enhancement coefficients on the order of 10$^6$
are primarily earned by long-range near-field effects and are independent of the effects from
hot-spots localized in the narrow gaps between
adjacent particles. Although the exact origin of the characteristic
length scale of 30\,nm is not clear now, the fact that it far
exceeds the mean radius of silver nanoislands implies that the
nature of field enhancement resides in the collective plasma response of
{\it particle arrays} and not in individual particles. The length scale is presumably related to the spatial spread (or depth of
penetration to the dielectric) of collective surface plasmon
polaritons. To clarify this, additional research that inspects the
SERS enhancement of nanoparticle agglomerations of different sizes
is needed.

By employing the weak distance dependence of SERS-enhancement of
2D nanoparticle arrays up to distances ~30\,nm, a new type of
SERS-substrates  might be developed with the passivating layer of
chemically passive and optically transparent dielectric deposited
on top of metal nanostructures of any quasi-2D morphology.
Virtually all key enhancement properties of a SERS substrate
should be preserved, yielding a surface that is cleanable,
non-expendable, and free of air degradation. This method of
conservation may help extend the life of most effective and quite
expensive (e.g., nanolithographically produced) SERS substrates,
thus making them more profitable.

Another application of long-range SERS enhancement lies in the
study of optical phenomena in low dimensional carrier systems
based on solid-state structures. This method is applicable even if
low dimensional objects are located at some depth-distance away
from the surface.

\begin{acknowledgments}
The authors acknowledge support from the Russian Foundation for
Basic Research.
\end{acknowledgments}


\begin{thebibliography}{99}

\bibitem{Fleischmann74} M. Fleischmann, P. J. Hendra and A. J. McQuillan, Chem.Phys.Lett. {\bf
26}(2), 163-166 (1974).

\bibitem{VanDuyne} D. L. Jeanmaire and R. P. Van Duyne, J. Electroanal.
Chem., {\bf 84}, 1-20 (1977).

\bibitem{Albrecht}  M. G. Albrecht and J. A. Creighton, Journal of the American Chemical Society,
{\bf 99}(15), 5215 (1977).

\bibitem{Kneipp1984}  K. Kneipp, D. Fassler, Chem. Phys. Lett., {\bf 106},
498 (1984).

\bibitem{Kneipp1997} K. Kneipp, Y. Wang, H. Kneipp, et al., Phys.Rev.Lett.
{\bf 78}, 1667 (1997).

\bibitem{GerstNitzan} J. I. Gersten, A. Nitzan, J. Chem. Phys. {\bf 75}, 1139
(1981).

\bibitem{Platzman80} S. L. McCall, P. M. Platzman, P. A. Wolff, Phys. Lett. A {\bf 77}, 381
(1980).

\bibitem{Wang80} D. -S. Wang, M. Kerker, H. W. Chew, Appl. Opt. {\bf 19}, 2315
(1980).

\bibitem{Moskov85} M. Moskovits, Rev. Mod. Phys. {\bf 57}, 783-826 (1985).

\bibitem{Otto84}  A. Otto, In Light scattering in solids IV. Electronic
scattering, spin effects, SERS and morphic effects;  M. Cardona,
G. Guntherodt, Eds.; Springer-Verlag: Berlin, Germany, 1984.

\bibitem{Otto92} A. Otto, I. Mrozek, H. Grabhorn, W. J. Akemann, J. Phys.
Condens. Matter {\bf 4}, 1143 (1992).

\bibitem{Kerker80}  M. Kerker, O. Siiman, L. A. Bumm, D. -S. Wang, Appl.
Opt. {\bf 19}, 3253 (1980).

\bibitem{Kerker81} D. -S. Wang, M. Kerker, Phys. Rev. B {\bf 24}, 1777 (1981).

\bibitem{Shalaev96} V. M. Shalaev, Phys. Rep. {\bf 272}, 61 (1996).

\bibitem{Sarychev98} V. M. Shalaev, A. K. Sarychev, Phys. Rev. B {\bf 57},
13265 (1998).

\bibitem{Moskov99}  V. A. Markel, V. M. Shalaev, P. Zhang, W. Huynh,
L. Tay, T. L. Haslett, M. Moskovits, Phys. Rev. B {\bf 59}, 10903
(1999).

\bibitem{Kennedy99}  B. J. Kennedy, S. Spaeth, M. Dickey, and K. T. Carron,
J. Phys. Chem. B {\bf 103}, 3640 (1999).

\bibitem{Compagnini99} G. Compagnini, C. Galati, and S. Pignataro, Phys. Chem. Chem. Phys., {\bf 1},
2351 (1999).

\bibitem{Faraday2006} J. A. Dieringer, A. D. McFarland, N. C. Shah, et al., Faraday
Discuss., {\bf 132}, 9 (2006).

\bibitem{Johnson72} P. B. Johnson and R. W. Christy, Phys. Rev.
B, {\bf 6}, 4370 (1972).

\end{thebibliography}
\end{document}